\documentclass[review,number,sort&compress,3p]{elsarticle}

\usepackage{amssymb}
\usepackage{amsmath}
\usepackage{amsbsy}
\usepackage{subfigure}
\usepackage{graphicx}
\usepackage{hyperref}

\begin{document}

\begin{frontmatter}

\title{Synthesis of pulses from particle detectors with a Generative Adversarial Network (GAN)}

\author[inta]{Alberto Regad\'io\corref{c1}}
\ead{regadioca@inta.es}

\author[wiyo]{Luis Esteban}
\ead{luis.esteban@die.upm.es}

\author[uahaut]{Sebasti\'an S\'anchez-Prieto}
\ead{sebastian.sanchez@uah.es}

\address[inta]{Department of Space Programs, Instituto Nacional de T\'ecnica Aeroespacial, 28850 Torrej\'on de Ardoz, Spain}
\address[wiyo]{WiYo Technologies, Cal\'endula 95, 28109 Alcobendas, Spain}
\address[uahaut]{Department of Computer Engineering, Space Research Group, Universidad de Alcal\'a, 28805 Alcal\'a de Henares, Spain}

\cortext[c1]{Corresponding Author}

\begin{abstract}
To address the possible lack or total absence of pulses from particle detectors during the development of its associate electronics, we propose a model that can generate them without losing the features of the real ones. This model is based on artificial neural networks, namely Generative Adversarial Networks (GAN). We describe the proposed network architecture, its training methodology and the approach to train the GAN with real pulses from a scintillator receiving radiation from sources of ${}^{137}$Cs and ${}^{22}$Na. The Generator was installed in a Xilinx's System-On-Chip (SoC). We show how the network is capable of generating pulses with the same shape as the real ones that even match the data distributions in the original pulse-height histogram data. 
\end{abstract}


\end{frontmatter}

\section{Introduction}
\label{Introduction}

The study of particles, both on the ground and in space, implies the use of detectors such as semiconductor, gas or scintillators. Access to the detectors and to their setup is not easy. In some cases it is required to include extra costs, long journeys to facilities, delays due to long waiting lists or even that the detectors are no longer operative. Finally, and depending on the detector, sometimes its events have a very low frequency, increasing the testing time. To avoid all this, creating realistic synthetic pulses may be a solution.

With this purpose we use Generative Adversarial Networks (GANs) \cite{Goodfellow2014}. They are a model based on deep neural networks to create new signals that mimic the original ones. Generic GANs consist of two main neural networks: the Generator and the Discriminator. The Generator {must be} specialized in creating data to fool the Discriminator into accepting it. Its adversary, the Discriminator, attempts to distinguish between samples drawn from the training data and samples drawn from the Generator. Both iterate in a balanced fight until the model reaches a possible point of equilibrium where it can generate convincing data and further iterations stop improving the model quality. This type of nets are difficult to train, namely because there is no learning curve as in other models, but rather a trade-off between Generator and Discriminator, as it will be seen in Section \ref{Results}. Furthermore, its use raises many open questions \cite{Odena2017} such as what kind of distributions can GANs model.

Although GANs may be composed entirely of fully connected layers, perhaps the most popular are the Deep Convolutional Generative Adversarial Network (DCGAN). They are an extension of the GANs, except that they use convolutional and convolutional-transpose layers in the Discriminator and Generator, respectively \cite{Radford2016}.

Much of the work related to DCGAN has focused on image style transfer, image filtering and image generation (e.g. CycleGAN \cite{Zhu2017}, StyleGAN \cite{Karras2019}, Alias-free GAN \cite{Karras2021}) with success. This image processing has been the starting point when using GAN networks in other types of applications, such as 1D signal processing. The change from 1D (time signals) to 2D (images) is achieved by generating spectrograms or scalograms with the input signal applying a Short-Time Fourier Transform (STFT) \cite{Smith2021}. This change of dimensions is not specific to GANs, it is also performed on other topologies such as U-Net (e.g. \cite{Choi2018}). The advantage of this approach is that neural networks developed for image processing can be reused to process other types of data.

However, there are also studies where signals are treated for what they really are: temporal sequences. In these works, the GAN models have included 1D convolutional networks \cite{Ramponi2018}, recurrent neural networks (RNN) \cite{Mogren2016} or Long-Short-Term Memory (LSTM) \cite{Odena2017, Esteban2017}. Some of these models also combine 1D layers with 2D layers.

GANs and their variations have been used extensively in detector simulation, mainly to perform simulations faster than other more computationally expensive simulators such as GEANT4 \cite{Allison2006} or PYTHIA8 \cite{Sjostrand2008}.

In particular, fully-connected GANs have been used to simulate Cherenkov detectors \cite{Derkach2020, Maevskiy2020} and for simulation of muons produced at the SHiP experiment \cite{Ahdida2019}. However, the most widely used GAN networks have been used to simulate component readouts in calorimeters, as in the case of \cite{Paganini2018, Paganini2018c, Buhmann2021}. Similar simulations have also been carried out for large calorimeters such as Atlas at the Large Hadron Collider (LHC) \cite{Salamani2018} at CERN. Most of these works have in common that they train the GANs to return voxelated images representing detectors and events like those generated by GEANT4.

In the field of physics, the simulation and reconstruction of single isolated particles produced in high energy collisions in electromagnetic and hadronic calorimeters has also been carried out with GANs \cite{Belayneh2020}, also taking as reference simulations with GEANT4.

GANs are also used for unfolding energy histograms \cite{Datta2018} in order to make the results independent of the details of the detectors. In the case of the aforementioned work, the GAN acts on the histogram taking as a reference the data obtained from PYTHIA8. In \cite{Bellagente2020} unfolding is carried out using Delphes3 \cite{DeFavereau2014} to generate the training data.

The enumerated works are aimed at calorimeters for large detectors, such as ATLAS at the LHC, where a large number of events occur, such as those generated by particle jets. However, these works do not focus on individual pulses nor on signals from the readout electronics of particle detectors.

On the contrary, the objective of this work is to create realistic pulses from a series of samples obtained from a real detector using GANs for the calibration of the readout electronics. It is important to remark that in the field of pulse spectroscopy, the pulse height is crucial to analyze and classify the incident radiation because it is proportional to its energy \cite{Knoll2010}. Besides, pulse shape varies depending on the type of detector used but it can also be analyzed to extract additional information about the particle type such as alpha or gamma for example. (see e.g. \cite{Knoll2010, Pausch1992, Pausch1994}).

For this purpose, it must be considered that mimicking the pulses is not enough. The pulse height histograms of synthetic pulses must also mimic those of real pulses so that the electronics tested with the proposed GAN can generate similar pulse height histograms to those that would be obtained with a real particle detector and their associated amplifier. This last objective brings up the question that if the data generated by GANs could add more events with statistical precision beyond the training data \cite{Matchev2020, Butter2020, Hao2018, Hao2020}.

This work has been evaluated for a scintillator detector but could be extrapolated to any type of detector adapting the parameters of the neural network presented in this article.

The rest of the manuscript is structured as follows: Section \ref{Model design} explains the GAN architecture, including both the Generator and the Discriminator, and their training process. The results of using these {neural networks} with simulated and real pulses are exposed in Section \ref{Results}. We conclude our work in Section \ref{Conclusions}.

%%%%%%%%%%%%%%%%%%%%%%%%%%%%%%%%%%%%%%%%%%%%%%%%%%%%%%%%%%%%%%%%%%%%%%%%%%%%%%%

\section{Model design}
\label{Model design}

\begin{figure}[!ht]
	\centering
	\includegraphics[scale=0.8]{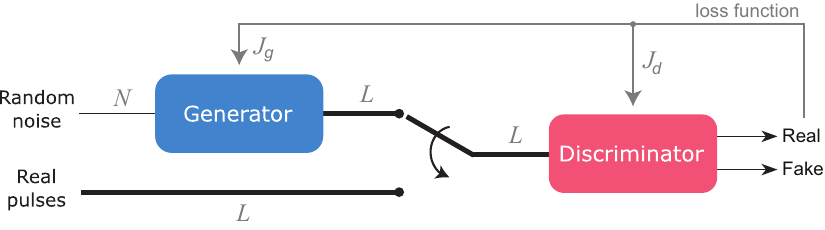}
	\caption{Architecture of the proposed GAN. $N$ is the dimension of the random vectors and $L$ is the pulse length.}
	\label{fig:Overview}
\end{figure}

A general scheme of the proposed GAN\footnote{Our code and data are available at github.com/arc140181/gan2pulses.} is depicted in Figure \ref{fig:Overview}. The details of these neural networks topology and a detailed description of their training process can be found in \cite{Goodfellow2014, Goodfellow2016b}. Basically, it consists of three processes that alternate for each batch: firstly, the Generator creates fake pulses from random values. Secondly, real and fake input pulses are mixed and the Discriminator is trained to distinguish them with supervised learning. Finally, the Generator weights are adjusted using backpropagation depending on the results that the Discriminator has obtained in the detection phase. Note that the generated pulses should also have a signal-to-noise ratio similar to the real pulses. The entire learning process is performed via backpropagation using the methodology taken from the original paper \cite{Goodfellow2014}, the loss functions are defined in the following sections.

We have chosen to use a DCGAN of 1D convolutional layers instead of time-sequence networks such as LSTM or RNN because we work with pulses contained in an array of constant length. With this structure, it is not necessary to do backpropagation through time, accelerating the training process. Moreover, neurons in one layer do not connect to all the neurons in the next layer, saving resources and training time. For simplicity, we will refer to the proposed network as GAN instead DCGAN from now on.

\subsection{Generator architecture}
\label{Generator architecture}

\begin{figure}[!ht]
	\centering
	\includegraphics[scale=0.7]{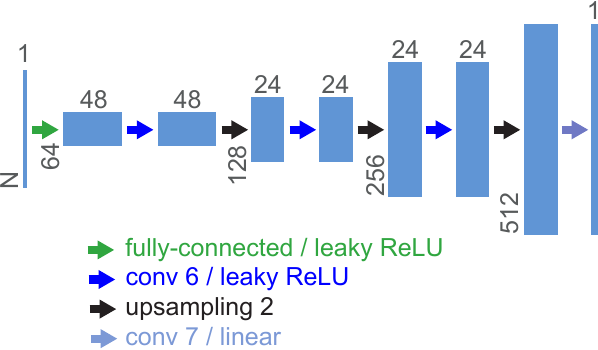}
	\caption{Generator architecture. The arrows stand for layers; on the left of the bar the type of layer is shown and on the right the activation function; `conv' stands for convolutional layer of kernel size equal the number that accompanies it. The rectangles represent the layer input/output matrices together with their dimensions.}
	\label{fig:Gen}
\end{figure}

As stated in Section \ref{Introduction}, the Generator is a neural network that takes a random vector $\mathbf{z}_i$ of dimension $N$ as input and returns realistic pulses (represented by $\mathbf{x}_i$). As it will be seen in the next section, a side effect of this is that the set of pulses generated also mimics the distribution of heights. At the beginning of the training process, it obviously generates noise at the output, but when {it} is trained, this noise is transformed into pulses. A diagram of the Generator used in this work is depicted in Figure \ref{fig:Gen}.

In the same way that original cost functions for GAN \cite{Goodfellow2014}, the loss functions of both Generator and Discriminator are calculated using the cross-entropy because it has offered better {evidence} for this specific task. Thus, the loss function of the Generator $J_g$ is defined as the function of the probability that its synthesized pulses have been labeled as fake by the Discriminator:

\begin{equation}\label{EqJg}	
	J_g = \frac{-1}{B} \sum_{i=0}^{B} \log D(G(\mathbf{z}_i))
\end{equation}
where $B$ is the batch size. $G: \mathbf{z}_i \rightarrow \mathbf{x}_i$ is the Generator's process, which implies that $G(\mathbf{z}_i)$ is a fake pulse. The Discriminator's process $D: \mathbf{x}_i \rightarrow [0,1]$ reflects its level of belief and goes from 0 (when an individual input pulse is classified as fake) to 1 (when is classified as true). Intuitively, whether the Generator provides pulses more and more similar to the real ones, the Discriminator has a hard task to distinguish them, thus assigning the probability that the pulse is real close to $D(G(\mathbf{z}_i))=0.5$.

\subsection{Discriminator architecture}
\label{Discriminator architecture}

\begin{figure}[!ht]
	\centering
	\includegraphics[scale=0.7]{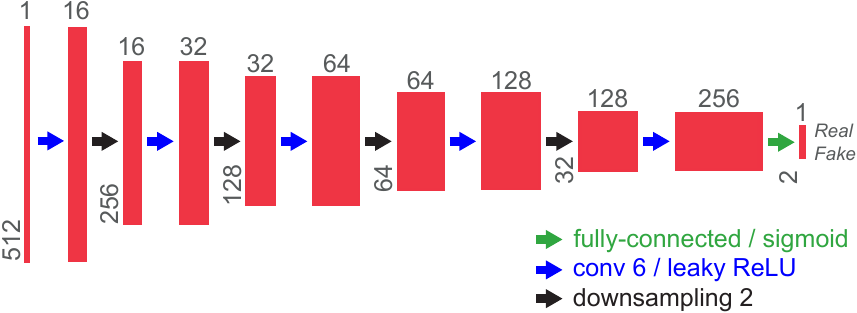}
	\caption{Discriminator architecture. The arrows stand for layers; on the left of the bar the type of layer is shown and on the right the activation function; 'conv' stands for convolutional layer of kernel size equal the number that accompanies it. The rectangles represent the layer input/output matrices together with their dimensions.}
	\label{fig:Discr}
\end{figure}

The Discriminator is the antagonistic part of the Generator. For this work, it has the architecture shown in Figure \ref{fig:Discr}. Except for the down-sampling layers, in the rest of the Discriminator they are of the same type as those of the Generator, although with different parameters.

The loss function of the Discriminator is

\begin{equation}\label{EqJd}	
	J_d = \frac{-1}{B} \sum_{i=0}^{B} \frac{1}{2} \log \left( D(\mathbf{x}_i) \right) + \frac{1}{2} \log \left(1 - D(G(\mathbf{z}_i)) \right)
\end{equation}

Ideally, a mean value of $D(\mathbf{x}_i) = D(G(\mathbf{z}_i)) = 0.5$ yields $J_d = 0.693$ which means that the Discriminator is not able to distinguish between real and fake pulses. This is the value to which $J_d$ {has} to tend for a successful training.

\subsection{Evaluation metrics}
\label{Evaluation metrics}

There are several methods to measure the quality of the individual pulses generated by a GAN. In some works related to time series such as \cite{Zhu2019}, both the Fr\'echet and Euclidean distances are used, but this measurement depends on the nature of the generated signal. In our case, it is difficult to measure the distance between one pulse and another due to the difference in shapes and heights inherent between them. For this reason, we decided to compare distances of real and synthetic pulses with similar heights.

As a complementary measure, we have decided to evaluate the pulses collectively using their pulse height histograms. Unfortunately, there is also no agreement as to how one should evaluate the quality of pulse height histograms generated with GANs. Sometimes, in the literature \cite{Delaney2019, Zhu2019} the data is evaluated using maximum mean discrepancy and a classical time series evaluation metric called dynamic time warping. In order to measure the distance between histograms, the Chi-squared or Wassertein metrics are sometimes used. However, in particle detector spectroscopy a common method to measure the quality of generated pulse-height histograms is the Full Width at Half Maximum (FWHM), defined as the width of the pulse height histogram at a level that is just half the maximum ordinate of the peak \cite{Knoll2010}. The FWHM indicates the resolution of the spectrometer, so it will also be used to measure the results of this work. In addition, 2D-histograms with the height and the negative amplitude of the pulse tail were generated to be compared to real histograms. Therefore, distances between pulses, distances between histograms, FHWM and 2D-histograms are measured in this work to assess the quality of the synthetic pulses.

%%%%%%%%%%%%%%%%%%%%%%%%%%%%%%%%%%%%%%%%%%%%%%%%%%%%%%%%%%%%%%%%%%%%%%%%%%%%%%%

\begin{figure*}[!htb]
	\centering
	\includegraphics[scale=0.8]{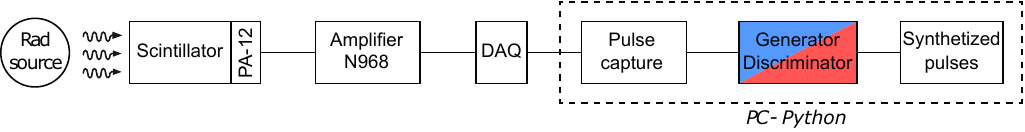}
	\caption{Diagram of the detection chain used for the experimental test. PM stands for photomultiplier and PA-12 is the preamplifier.}
	\label{fig:DetectionChain}
\end{figure*}

%%%%%%%%%%%%%%%%%%%%%%%%%%%%%%%%%%%%%%%%%%%%%%%%%%%%%%%%%%%%%%%%%%%%%%%%%%%%%%%

\section{Results}
\label{Results}

In order to evaluate the proposed approach, a set of tests were performed. The implementation of the GAN was programmed in Python using Tensorflow 2 \cite{TENSORFLOW2015} and Keras \cite{KERAS2015} packages together with their \texttt{Conv1D} function to implement 1D convolutional layers. The number of parameters of the Generator and the Discriminator are of 845,329 and 265,809, respectively. This number of parameters and their ratio is the best we have found so far in terms of the trade-off between net complexity and its results. The pulse height to create the histograms was measured using the \texttt{peak} function and the FWHM measurement was carried out using the function \texttt{peak\_widths} with \texttt{rel\_height = 0.5} in all cases. Both functions belong to the \texttt{scipy.signal} package \cite{SCIPY2020}.

The pulses to train the GAN were collected in the Radiation Physics Laboratory located in Santiago de Compostela University (Spain) using a scintillator. A diagram of the detection chain used in the experimental test is shown in Figure \ref{fig:DetectionChain}. The scintillator model of NaI is the 1M1/1.5 working at +475 V. It includes a phototube to amplify the optical signal, with an integrated preamplifier PA-12. The amplifier N968 (with a shaping of 2 $\mu$s and gain $\times$ 14 was connected to a Digital Phosphor Oscilloscope Tektronix TDS 3014B. An amount of {L = 512} points were taken for each pulse at a frequency of 1 {GHz}. {These pulses have been digitally filtered with a FIR filter whose transfer function is} $h[z] = \frac{1}{4}(1+{z}^{-1}+{z}^{-2}+{z}^{-3})$. Therefore, the shape and heights of observed pulses are formed by these stages.

The oscilloscope is used as a data acquisition system, receiving raw data from the amplifier and storing them in a laptop. The resolution of the signal amplitude is 256 bits for a range between $-$5 and 5 V. {We work with 8-bit signed integer as input.} The scintillator received radiation from a source of ${}^{137}$Cs whose activity is 8.71 kBq and produces peaks at 661.6 keV. {The raw data was stored in files to allow data reuse without recapturing new samples, allowing reusability and ensuring that changes in the results obtained during the tests are exclusively due to modifications of the GAN parameters.}

In order to train the GAN, a set {5,000} different pulses captured from this facility were used. As mentioned above, each pulse has a length {$L = 512$} samples. The input signal of the {G}enerator were random vectors of dimension $N$ from a normal distribution with zero mean and variance equal to one. This same type of distribution was used to generate the pulses when the training stage was finished. The chosen dimension was $N=128$. Values of $N<10$ make the pulses noiseless and therefore unrealistic, but a correct histogram can be generated with them. {Values of $N$ beyond 128 increase significantly the time of training.}

The chosen batch dimension was $B=128$ although values up to 256 were tested without significant modifications in the results. When $B$ tends to 1, the pulse shape is mimicked in a shorter period of epochs but the resulting histograms barely show the characteristic peaks of the radiation sources. The learning algorithm was \texttt{Adam} adjusted to \texttt{learning\_rate=0.00004} and \texttt{beta1=0.5} for both Generator and Discriminator. The training stops when the FWHM of the histogram with synthetic pulses is equal to the FWHM of the real pulses or when 400 epochs are reached. The training process took 29 minutes for 238 epochs on Google Colab with its Graphics Processing Unit (GPU) enabled.

Once trained, the Generator was introduced in the evaluation board Pynq-Z2 (Python Productivity for Zynq) that hosts a Programmable System-on-Chip (SoC) based on an Field Programmable Gate Array (FPGA) \cite{PYNQ2022}, a Xilinx Zynq Z7020 with two ARM Cortex-A9. The Generator has been executed using the Numpy 1.13.0 and Tensorflow 1.1.0 libraries. The low version of Tensorflow and the limited processing capacity of this SoC are the reasons why we have chosen to train previously the model with Google Colab as explained. Once the weights have been introduced, we have the pulses of a particle detector embedded in a portable board that also owns enough reconfigurable logic to perform the desired digital processing of them. 

\begin{figure}[!htb]
	\centering
	\includegraphics[scale=0.8]{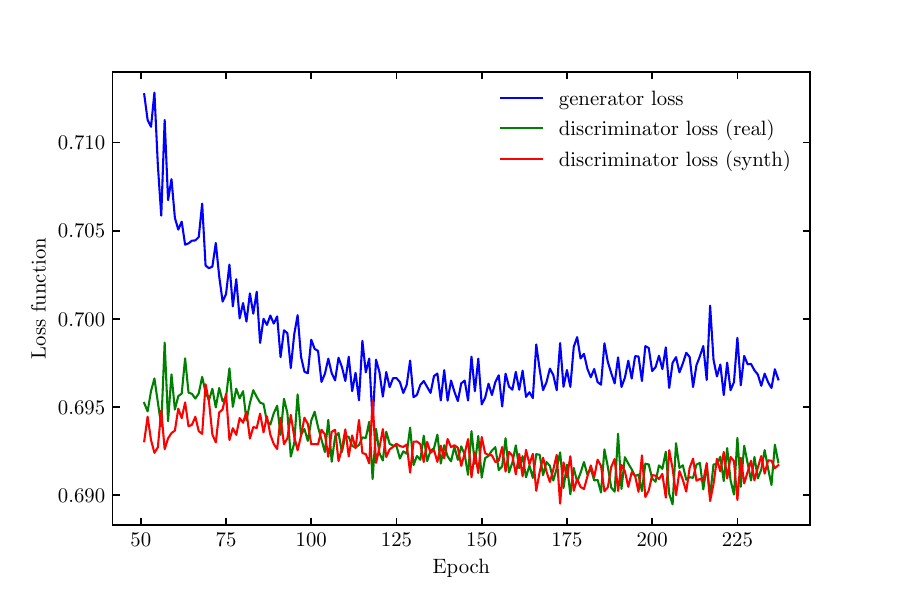}
	\caption{Learning curve for ${}^{137}$Cs. The y-axis stands for loss functions according to Eq. (\ref{EqJg}) and (\ref{EqJd}) for the Generator and the Discriminator, respectively. Note that both functions of the discriminator are overlapping.}
	\label{fig:LearningCurve}
\end{figure}

Figure \ref{fig:LearningCurve} shows the loss function of the Generator and the Discriminator, defined as explained in Sections \ref{Generator architecture} and \ref{Discriminator architecture}, respectively. {As it can be observed, the three functions are balanced throughout the entire training stage. This figure also shows a line plot of the Discriminator loss function on real and fake pulses during training that overlaps. This overlapping indicates that the Discriminator fails the same with both real and generated pulses, so it can be concluded that the Generator creates plausible pulses.}

\begin{figure*}[!hbt]
	\centering
	\includegraphics[scale=0.7]{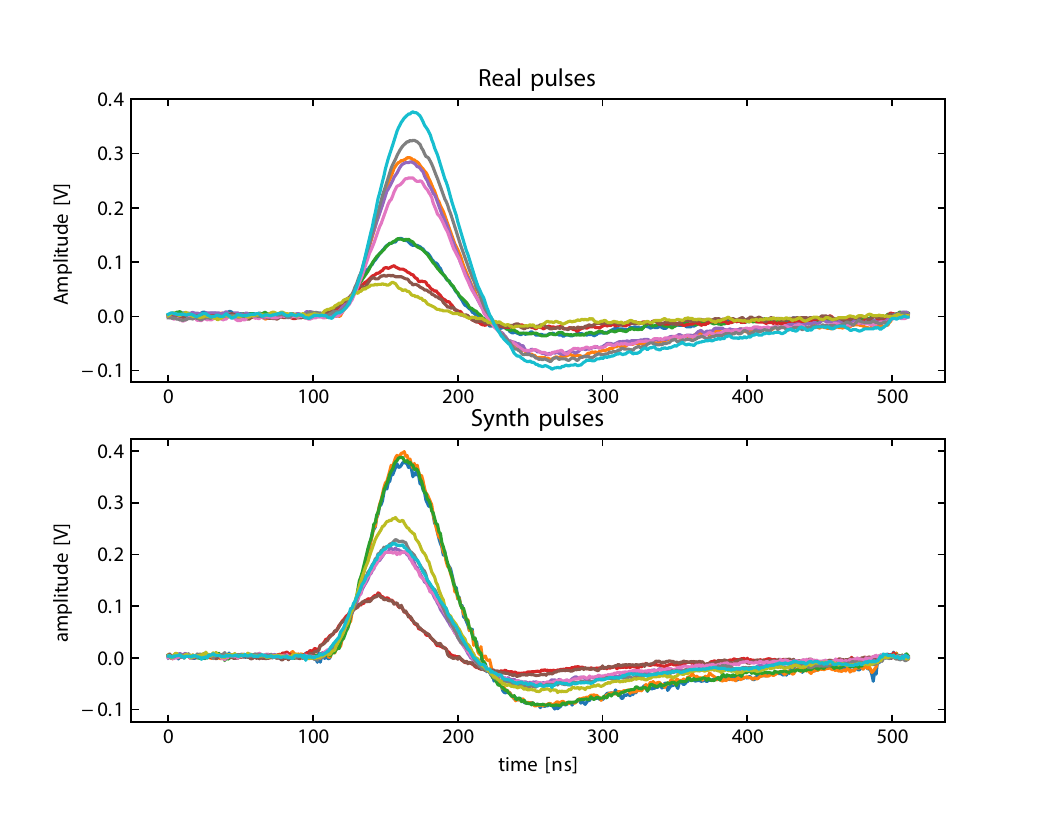}
	\caption{Pulse shape comparison in time domain for ${}^{137}$Cs source. {The colors are only used to better distinguish the pulses from each other.}}
	\label{fig:ExPulsesCs}
\end{figure*}

In Figure \ref{fig:ExPulsesCs} a comparison between 10 synthetic pulses and 10 real pulses from the described scintillator and the ${}^{137}$Cs source is shown. At first glance they seem similar.

\begin{figure}[!htb]
	\centering
	\includegraphics[scale=0.8]{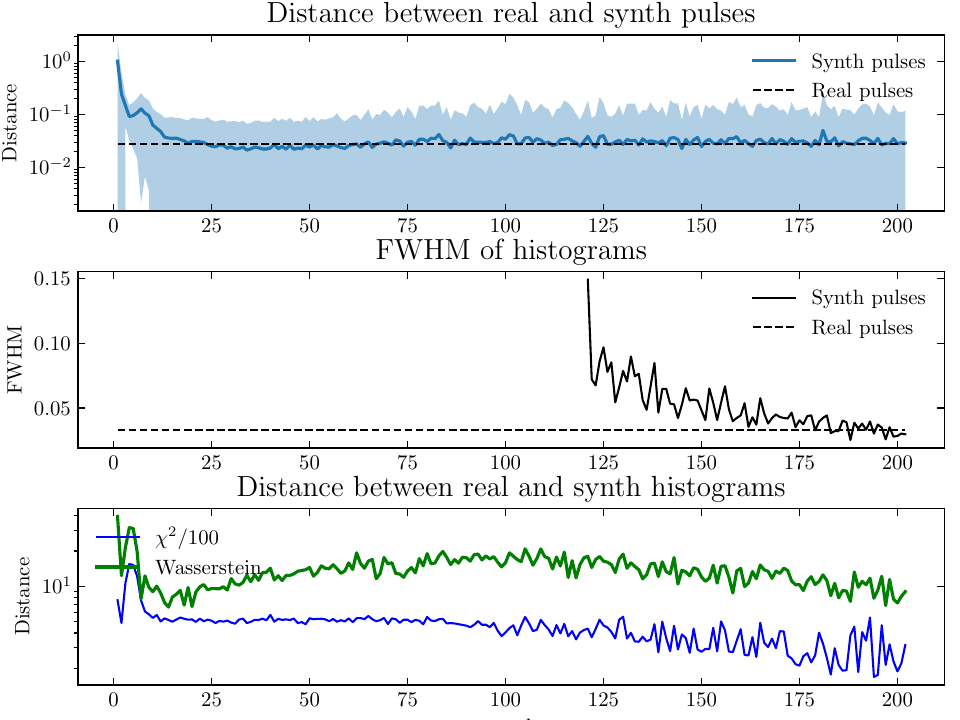}
	\caption{Training process for GAN using ${}^{137}$Cs pulses. Top: Mean and standard deviation of the quadratic distance between real and synthetized pulses. Center: Wasserstein and Chi-squared distances for histograms. Bottom: FWHM of the histograms.}
	\label{fig:FWHMDistanceCs}
\end{figure}

Figure \ref{fig:FWHMDistanceCs} shows the results of the training progresses according to the parameters listed in Section \ref{Evaluation metrics}. Specifically, the figure above shows the quadratic distance between the real and synthesized pulses, the central figure shows the distances between histograms and the figure below shows the FWHM. The latter depicts how the FWHM tends to the value given by the real histogram as the training process progresses. Note that the FWHM in the early epochs is not shown because the histogram in early training stages is mostly flat.

We therefore observe that during training the first thing that is achieved is to mimic the shape of the pulses, something logical since it is what is intended with the loss functions (\ref{EqJg}, \ref{EqJd}). In addition, the GAN network also mimics the shape of the histograms as training progresses.

\begin{figure}[!htb]
	\centering
	\includegraphics[scale=0.8]{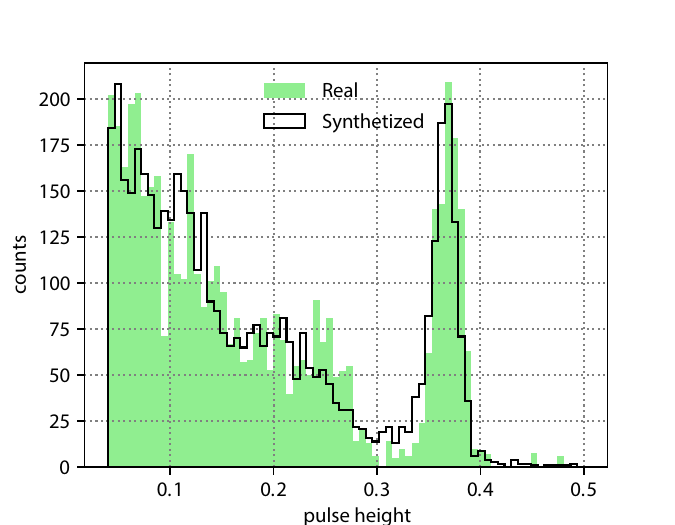}
	\caption{Energy histograms for ${}^{137}$Cs {generated with 5,000 real pulses and 5,000 synthetic pulses}.}
	\label{fig:HistCs}
\end{figure}

\begin{figure}[!htb]
	\centering
	\includegraphics[scale=0.8]{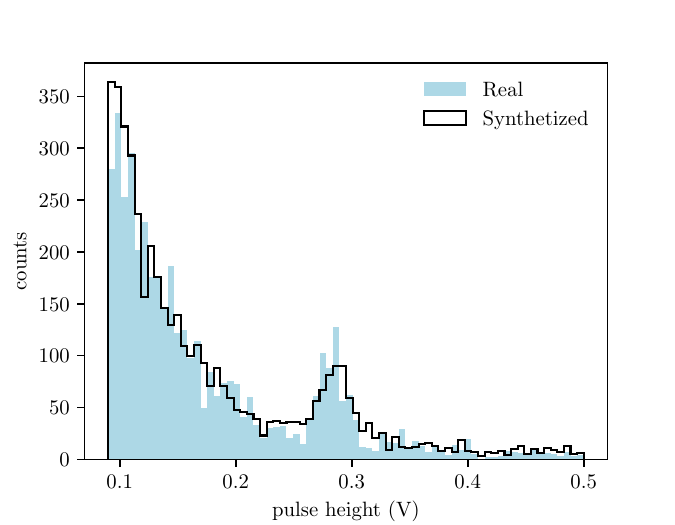}
	\caption{Energy histograms for ${}^{22}$Na {generated with 5,000 real pulses and 5,000 synthetic pulses}.}
	\label{fig:HistNa}
\end{figure}

Figure \ref{fig:HistCs} shows the histogram generated with 5,000 pulses from the described scintillator and the source of ${}^{137}$Cs, overlapped with another histogram generated with 5,000 synthetic pulses. In this histogram and the one shown in Figure \ref{fig:HistNa}, a pulse height equal to 1 is equivalent to 1760 keV. It can be seen that both histograms are similar throughout their entire spectrum.

\begin{figure}[!htb]
	\centering
	\includegraphics[scale=0.8]{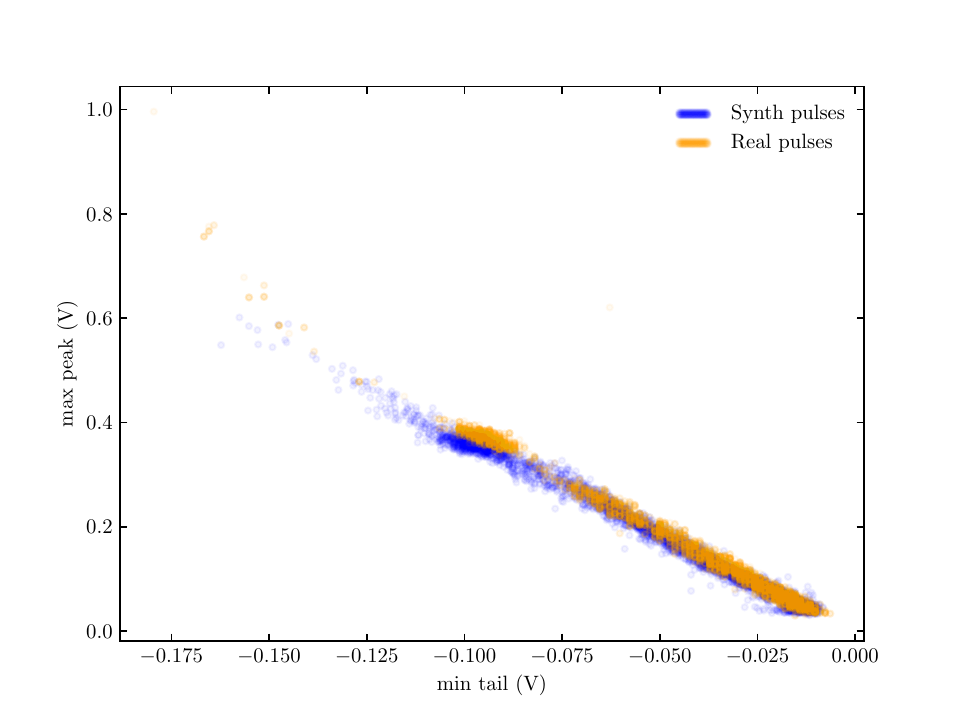}
	\caption{Two-dimensional energy histograms for ${}^{137}$Cs {generated with 5,000 real pulses (top) and 5,000 synthetic pulses (bottom)}.}
	\label{fig:Hist2DCs}
\end{figure}

Finally, in order to check the pulse shape, we generated a 2D histogram with the pulse height and the undershoot amplitude (Figure \ref{fig:Hist2DCs}). According to Figure \ref{fig:ExPulsesCs}, the undershoots occur at about 250 $\mu$s. We observe that in both cases, maximum and minimum values of real and synthetized pulses are similar.

\begin{figure}[!htb]
	\centering
	\includegraphics[scale=0.8]{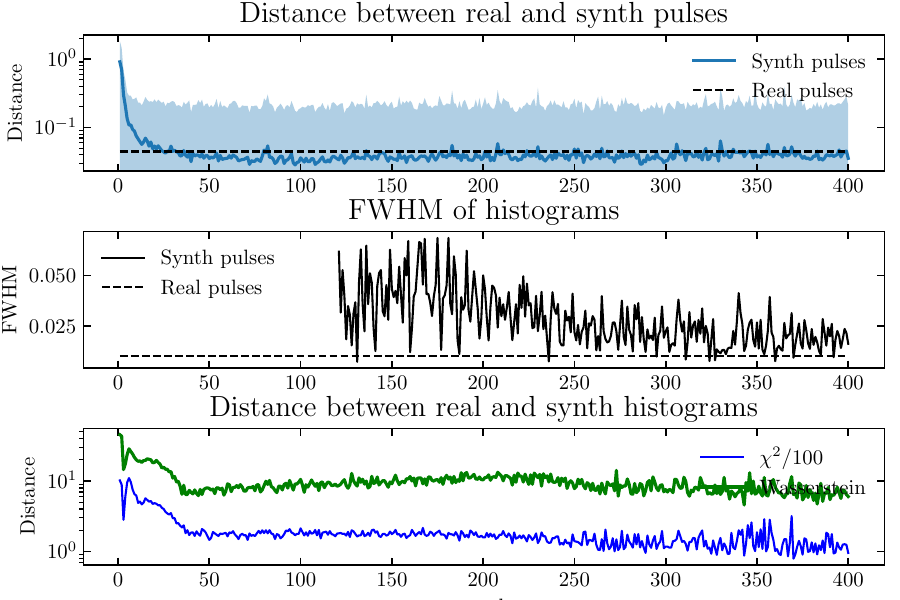}
	\caption{Training process for GAN using ${}^{22}$Na pulses. Top: Mean and standard deviation of the quadratic distance between real and synthetized pulses. Center: Wasserstein and Chi-squared distances for histograms. Bottom: FWHM of the histograms}
	\label{fig:FWHMDistanceNa}
\end{figure}

\begin{figure}[!htb]
	\centering
	\includegraphics[scale=0.8]{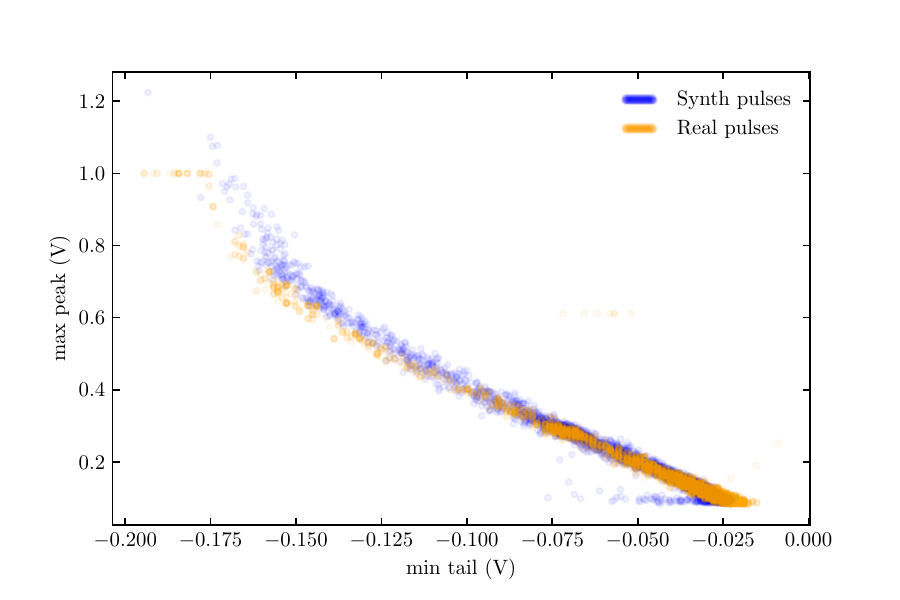}
	\caption{Two-dimensional energy histograms for ${}^{22}$Na {generated with 5,000 real pulses (top) and 5,000 synthetic pulses (bottom)}.}
	\label{fig:Hist2DNa}
\end{figure}

Another test was performed substituting ${}^{137}$Cs by a source of ${}^{22}$Na whose activity is 105 kBq and produces peaks at 511 keV with similar results. With identical Generator and Discriminator parameters, we can see in Figure \ref{fig:FWHMDistanceNa} that the FWHM also converges to that of the real histogram despite the height of its peak is almost half of the one shown in Figure \ref{fig:FWHMDistanceCs}. Moreover, in Figure \ref{fig:HistNa} we can see how both generated and real histograms are also similar. Finally, Figure \ref{fig:Hist2DNa} shows similar results in pulse height and undershoot amplitude for the same source.

It is worth to mention that in preliminary tests, apart from the cost functions of Eq. (\ref{EqJg}, \ref{EqJd}), an additional cost function that take into account the distance between real and generated histograms was added. The problem with this new cost function is that it was computationally expensive and therefore it was discarded, reducing the fitting of real and generated histograms. However, from a conceptual point of view, it is important to note that only with the cost function of the mean distance between pulses, the GAN was able to also infer the histograms.

\begin{figure}[!htb]
	\centering
	\includegraphics[scale=0.8]{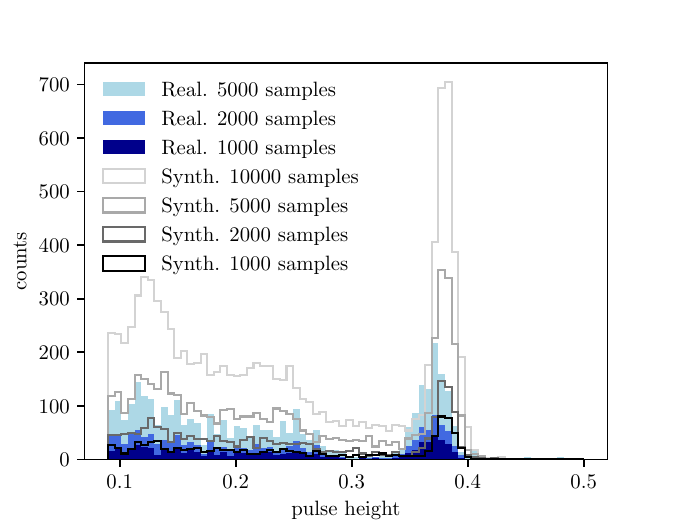}
	\caption{Histograms of ${}^{137}$Cs generated with real and synthetic pulses. The FWHM with 5,000 real pulses is 0.28 and the FWHM whit 10,000 synthetic pulses is 0.20.}
	\label{fig:HistMulCs}
\end{figure}

Finally, an overlay of all histograms is shown in Figure \ref{fig:HistMulCs}. In this figure, it is observed that as the synthesized pulses increase, the width of the main peak, and therefore the FWHM, decreases. It is also observed that the comb structure that sometimes appears in certain parts of the histogram is gradually eliminated. All this shows that the GAN learns the general structure of the histogram.

%It is observed that when the number of synthesized pulses is increased, the width of the main peak, and therefore the FWHM, is decreased. It is also observed that the comb structure that sometimes appears in certain parts of the histogram is gradually eliminated. All this shows that the GAN learns the general structure of the histogram, improving its resolution and making it better to use it than the 5,000 real pulses.

To summarize, it is therefore expected that a GAN topology equal to that shown in Figure \ref{fig:Gen} and \ref{fig:Discr}, together with the cost functions defined in Eq. (\ref{EqJg}, \ref{EqJd}) can work with a detection chain similar to that Figure \ref{fig:DetectionChain} simply by varying parameters such as number of layers, input dimensions or learning rate.

%%%%%%%%%%%%%%%%%%%%%%%%%%%%%%%%%%%%%%%%%%%%%%%%%%%%%%%%%%%%%%%%%%%%%%%%%%%%%%%

\section{Conclusions}
\label{Conclusions}

We present a novel model that can generate pulses that mimic the features of real pulses from readout electronics of particle detectors. This approach is based on GANs. After training with real pulses from a scintillator, our model can create synthetic pulses that are similar to the real ones. This similarity has been measured by calculating the distances between individual pulses and by means of the distance and the FWHM of the histograms that they generate. The main utility of this GAN is to significantly accelerate the development of electronics associated with particle detectors by providing the desired number of pulses, at the desired arrival rate. The Generator was installed in a SoC based on a Xilinx Zynq Z7020 that includes two ARM Cortex-A9 processors with Python installed. Therefore, leveraging the usage of detectors, which are now replaced by a portable board that also has reconfigurable logic to implement the digital signal processing of the pulses. This board allows a first phase of development of the readout electronics and pulse processing without a physical detector nor the associated electronic equipment and therefore dramatically reducing the testing costs.

\section*{Acknowledgement}

The authors thank the Radiation Physics Laboratory located in Santiago de Compostela University (Spain) for providing the scintillator detector and its associated electronics. %{The authors also want to thank the anonymous reviewers for helping to greatly improve the quality of this article.}

\bibliography{thebib}

\end{document}